\documentclass[twocolumn,showpacs,showkeys,preprintnumbers,superscriptaddress,amsmath,floatfix,amssymb,secnumarabic, nofootinbib]{revtex4}
\usepackage[colorlinks=true]{hyperref}
\usepackage{graphicx}
\usepackage{epsfig}

\def\({\left(}
\def\){\right)}
\def\[{\left[}
\def\]{\right]}

\newcommand{\lr}[1]{ \left( #1 \right) }
\newcommand{\lrs}[1]{ \left[ #1 \right] }

\newcommand{\vev}[1]{ \langle \, #1 \, \rangle }

\newcommand{\expa}[1]{ \exp{\left( #1 \right)} }

\newcommand{\comment}[1]{}

\newcommand{\logo}{\\ \vskip -18mm
\leftline{\includegraphics[scale=0.3,clip=false]{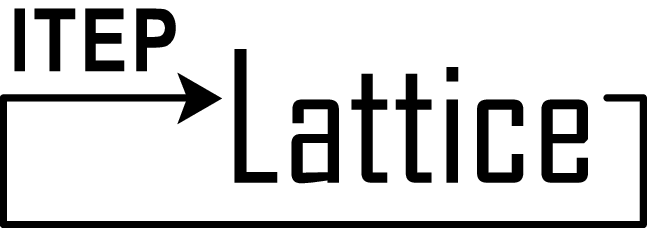}} \vskip 10mm}

\begin{document}
\sloppy

\title{Numerical simulation of graphene in external magnetic field \logo}

\author{D.~L.~Boyda}
\email{boyda_d@mail.ru}
\affiliation{Far Eastern Federal University, School of Biomedicine, Vladivostok, 690091 Russia}

\author{V.~V.~Braguta}
\email{braguta@mail.ru}
\affiliation{Institute for High Energy Physics, Protvino, 142281 Russia}
\affiliation{Institute of Theoretical and Experimental Physics, Moscow, 117218 Russia}

\author{S.~N.~Valgushev}
\email{semuon06@gmail.com}
\affiliation{Institute of Theoretical and Experimental Physics, Moscow, 117218 Russia}
\affiliation{Moscow Institute for Physics and Technology, Dolgoprudny, 141700 Russia}

\author{M.~I.~Polikarpov}
\affiliation{Institute of Theoretical and Experimental Physics, Moscow, 117218 Russia}
\affiliation{Moscow Institute for Physics and Technology, Dolgoprudny, 141700 Russia}

\author{M.~V.~Ulybyshev}
\email{ulybyshev@goa.bog.msu.ru}
\affiliation{Institute of Theoretical and Experimental Physics, Moscow, 117218 Russia}
\affiliation{Institute for Theoretical Problems of Microphysics, Moscow State University, Moscow, 119899 Russia}


\begin{abstract}

In this paper the results of numerical simulation of monolayer graphene in external magnetic field are presented.
The numerical simulation is performed in the effective lattice field theory with noncompact $3 + 1$-dimensional
Abelian lattice gauge fields and $2 + 1$-dimensional staggered lattice fermions. The dependences of fermion
condensate and graphene conductivity on the dielectric permittivity of substrate for different values of external
magnetic field are calculated. It is found that magnetic field shifts insulator-semimetal phase transition
to larger values of the dielectric permittivity of substrate. The phase diagram of graphene in external
magnetic field is drawn.

\end{abstract}

\maketitle

\section{Introduction}
\label{IntroductionSec}

Graphene is a two dimensional crystal composed of carbon atoms packed in a honeycomb lattice.
This material is well known due to its low energy electronic spectrum which can be described
by an effective theory with two massless Dirac fermions living in two dimensions
\cite{Novoselov:04:1, Novoselov:09:1, Geim:07:1, semenoff}. Due to this striking property
graphene is a perspective material both for study of different relativistic quantum
field theory phenomena and development of new electronic devices.

We base on the effective field theory of graphene.
Within this approximation the massless electronic excitations in graphene propagate with the
speed $v_F \simeq c/300$. Since this speed is much smaller than the speed of light--$c$
the interaction between quasiparticles in graphene can be approximated by instantaneous
Coulomb law with the effective coupling constant $\alpha_{eff}\sim \alpha_{em} c /v_F \sim 300/137 \sim 2$.
From this fact one can conclude that electronic excitations in graphene form
strongly interacting many body system.

In real experiments graphene is put on substrate. The effective coupling constant
for graphene on substrate with the dielectric permittivity $\epsilon$
is smaller by a factor $2/(\epsilon+1)$ due to the screening. The variation of dielectric
permittivity $\epsilon$ of  substrate changes the effective coupling constant and thus allows
to study the properties of graphene in both strong and week coupling regime.

In weak coupling regime theoretical description of graphene properties based on perturbation
theory gives reliable results. In strong coupling regime there are no theoretical approaches based on first
principles. From this perspective Monte-Carlo simulation of graphene is the only possibility to study graphene
when interaction between quasiparticles is strong.

Recently there appeared a lot of papers where graphene electronic properties were studied numerically
\cite{Lahde:09:1, Lahde:09:2, Lahde:09:3, Drut:10:1, Lahde:11:1, Hands:08:1, Hands:10:1, Hands:11:1, Buividovich:2012uk, Buividovich:2012nx, Braguta:2013klm}.
  The results obtained in these papers reveal one picture of insulator-semimetal phase transition in graphene - it takes place at the dielectric permittivity $\epsilon \sim 4$. At weak coupling regime graphene is in the semimetal phase. In this phase the conductivity is $\sigma \sim e^2/h$ and
there is no gap in the spectrum of fermionic excitations. The chiral symmetry of graphene
is not broken. At strong coupling regime graphene is in the insulator phase. In this phase the conductivity is
considerably suppressed and there is an energy gap in in the spectrum of fermionic excitations.
The opening of the energy gap is accompanied by the appearance of the fermion
chiral condensate $\langle \bar \psi \psi \rangle$, which breaks chiral symmetry of graphene.

 But in the most recent paper \cite{Ulybyshev:2013swa} it was shown that modification of the quasiparticles interaction potential
at small distances dramatically changes the picture of phase transitions in graphene. Suspended graphene is still a conductor (in agreement with experiment) and phase transition is shifted into unphysical region $\epsilon < 1$.  In this paper
we don't make this modification and carry out the simulation as it was done in paper \cite{Buividovich:2012uk}. This simulations don't correspond to experimental results numerically but we still hope that the description of graphene phase diagram will be qualitatively reasonable.

The aim of this paper is to study graphene phase diagram in the external magnetic field perpendicular
to the graphene plane. To carry out this investigation the Monte-Carlo simulation
of graphene with dynamical staggered fermions \cite{Buividovich:2012uk}  will be applied.
This study is motivated by the
theoretical interest to understand how external magnetic field can change graphene properties. In addition,
the possibility to control graphene properties through the external magnetic field seems
very promising in the development of new electronic devices.

This paper is organized as follows. In the next section a brief review of the simulation algorithm is given.
In the last section the results of this paper are presented and discussed.

\section{ Lattice simulation of graphene }

\subsection{Simulation algorithm.}
\label{subsec:basic_defs}

The partition function of graphene can be written in the following form  \cite{Novoselov:04:1, Novoselov:09:1,
Geim:07:1, semenoff}
\begin{eqnarray}
\label{partfan}
 \mathcal{Z}   =
 \int \mathcal{D}\bar{\psi} \mathcal{D}\psi \mathcal{D}A_0
 \exp\left( -\frac{1}{2}\int d^4x \lr{\partial_{i} A_{0}}^2
 - \right. \nonumber\\ \left. -
 \int d^3x \, \bar{\psi}_f \, \lr{ \Gamma_0 \, \lr{\partial_0 - i g A_0}
 - \sum\limits_{i=1,2}
    \Gamma_i \partial_i} \psi_f
 \right) ,
\end{eqnarray}
where $A_{0}$ is the zero component of the vector potential of the $3 + 1$ electromagnetic field, $\Gamma_{\mu}$  are Euclidean gamma-matrices and
$\psi_f$ ($f = 1, 2$) are two flavours of Dirac fermions which correspond to two spin components of the non-relativistic electrons in graphene.
Effective coupling constant is $g^2 = 2 \alpha_{em}/( v_F (\epsilon+1))$ ($\hbar=c=1$ is assumed). It is worth to note that partition function (\ref{partfan}) doesn't contain dynamical vector part of the potential $A_i$, since the inclusion of this part leads to the corrections which are suppressed by the factor $v_F/c  \sim 1/300$.

Zero component  of the vector potential
$A_0$ satisfies periodic boundary condition in space and time $A_{0} (t=0)=A_0(t=1/T)$, where $T$ is temperature. In the absence of
magnetic field fermion spinors  satisfy periodic boundary condition in space and antiperiodic boundary condition in the time
direction $\psi_f(t=0)=-\psi_f(t=1/T)$.
If magnetic field is switched on, fermion boundary condition in space should be modified \cite{AlHashimi:2008hr}.

The simulation of partition function (\ref{partfan}) is carried out within the approach developed in  \cite{Buividovich:2012uk}.
In order to discretize the fermionic part of the action in (\ref{partfan}) the  staggered fermions \cite{MontvayMuenster, DeGrandDeTarLQCD} are used.
One flavor of staggered fermions in $2 + 1$   dimensions corresponds to two flavors of continuum Dirac fermions \cite{ MontvayMuenster, DeGrandDeTarLQCD, Burden:87:1},
which makes them especially suitable for the simulations of the graphene effective field theory.

 The action for staggered fermions coupled to Abelian lattice gauge field is
\begin{eqnarray}
\label{lat_act_interact}
S_{\Psi}\lrs{\bar{\Psi}_x, \Psi_x, \theta_{x, \, \mu}} =
 \sum\limits_{x, y} \bar{\Psi}_x \, D_{x, y}\lrs{\theta_{x, \, \mu}} \, \Psi_y
= \nonumber \\ =
 \frac{1}{2} \, \sum\limits_{x} \, \delta_{x_3, \, 0} \, \left(  \sum\limits_{\mu=0, 1, 2}
 \bar{\Psi}_x \alpha_{x, \mu} e^{i \theta_{x, \, \mu}} \Psi_{x+\hat{\mu}}
 - \right. \nonumber \\ \left. -
 \sum_{\mu=0, 1, 2}
 \bar{\Psi}_x \alpha_{x, \mu} e^{-i \theta_{x, \, \mu}} \Psi_{x-\hat{\mu}}
 + m {\bar{\Psi}}_x \Psi_x \right) ,
\end{eqnarray}
where the lattice coordinates $x$ take integer values $x^{\mu} = 0 \ldots L_{\mu}-1$ and $x^3$ is restricted to $x^3 = 0$,
 $\bar{\Psi}_x$ is a single-component Grassman-valued field, $\alpha_{x, \mu} = (-1)^{x_0 + \ldots + x_{\mu-1}}$, and $\theta_{x, \, \mu}$ are the link variables which are the lattice counterpart of the vector potential $A_{\mu}\lr{x}$. For further convenience, we have also introduced the matrix elements $D_{x, y}$ of the staggered Dirac operator.

It should be noted here that nonzero mass term in (\ref{lat_act_interact}) is necessary in order
to ensure the invertibility of the staggered Dirac operator $D_{x, y}$. Numerical results in the physical
  limit of zero mass is obtained by performing simulations at several nonzero values of $m$ and by extrapolating the expectation values of physical observables to $m \rightarrow 0$.

To discretize the electromagnetic part of partition function (\ref{partfan}) noncompact action is used
\begin{eqnarray}
\label{gauge_lat_act}
 S_g\lrs{\theta_{x, \, \mu}} = \frac{\beta}{2} \, \sum\limits_x \sum\limits^{3}_{i=1}
 \lr{ \theta_{x, \, 0} - \theta_{x + \hat{i}, \, 0} }^2  ,
\end{eqnarray}
where the summation is carried out over all lattice. The constant $\beta$ is defined as follows
\begin{eqnarray}
\label{lattice_coupling_constant}
 \beta \equiv \frac{1}{g^2} = \frac{v_F}{4 \pi e^2} \, \frac{\epsilon + 1}{2}.
\end{eqnarray}
The factor $\frac{\epsilon + 1}{2}$ takes into account the electrostatic screening of the Coulomb interaction due to the substrate.

  The introduction of nonzero homogeneous magnetic field $H$ perpendicular to graphene plane can be done in a standard way
through the modification of the link variable $\theta_{x, i}, i=1, 2$, which corresponds to the vector potential
$A_i=H (x_2 \delta_{i1}-x_1 \delta_{i2})/2$. As the result of torus geometry of the lattice the flux
of the magnetic field through the whole lattice $\Phi$ is quantized as
\begin{eqnarray}
\Phi=\frac {2 \pi} {e } n,
\label{quantum}
\end{eqnarray}
where $n$ is a digit number. Quantization of the flux $\Phi$ leads to the quantization of the magnetic field which
can be modeled on the lattice
\begin{eqnarray}
H=\frac {2 \pi} {e L_s^2} n,
\end{eqnarray}
where $L_s$ is the size of the lattice.

 Since action (\ref{lat_act_interact}) is bilinear in fermionic fields, they can be integrated out
\begin{eqnarray}
\label{ferm_integrated}
 \mathcal{Z} =
 \int \mathcal{D}\bar{\Psi}_x \, \mathcal{D}\Psi_x \, \mathcal{D}\theta_{x, \, 0}\,
 \nonumber \\
 \exp\lr{ - S_g\lrs{\theta_{x, \, 0}} - S_{\Psi}\lrs{\bar{\Psi}_x, \Psi_x, \theta_{x, \, 0}}}
 = \nonumber \\ =
 \int \mathcal{D}\theta_{x, \, 0}\,
 \det\lr{ D\lrs{ \theta_{x, \, 0} } }
 \expa{ -S_g\lrs{\theta_{x, \, 0}} }  .
\end{eqnarray}
Thus one gets the following expression
\begin{eqnarray}
\label{eff_action}
 S_{eff}\lrs{ \theta_{x, \, 0} } = S_g\lrs{ \theta_{x, \, 0} } - \ln \det\lr{ D\lrs{ \theta_{x, \, 0} } } ,
\end{eqnarray}
which includes the determinant $\det\lr{ D\lrs{ \theta_{x, \, \mu} }}$ of the Dirac operator $D_{x, y}\lrs{\theta_{x, \,
\mu}}$ introduced in (\ref{lat_act_interact}).

To generate configurations of the field $\theta_{x, \, 0}$
with the statistical weight $\expa{-S_{eff}\lrs{ \theta_{x, \, 0} }}$
the standard Hybrid Monte-Carlo Method is used \cite{MontvayMuenster, DeGrandDeTarLQCD, Lahde:09:2}.
 In order to speed up the simulations we also perform local heatbath updates of the gauge field outside of the graphene plane (at $x^3 \neq 0$) between Hybrid Monte-Carlo updates.
 Both algorithms satisfy the detailed balance condition for weight (\ref{ferm_integrated}) \cite{MontvayMuenster, DeGrandDeTarLQCD}. Successive application of these algorithms does not,
  in general, have this property. Nevertheless, by using the composition rule for transition probabilities it is easy to demonstrate that the path integral weight (\ref{ferm_integrated}) is still the stationary probability distribution for such a combination of both algorithms. While local heatbath updates are computationally very cheap, they significantly decrease the autocorrelation time of the algorithm.

\subsection{Physical observables on the lattice}
\label{subsec:observables}

 The main goal of this paper is to measure the electric conductivity of graphene in external magnetic field,
that is, a linear response of the electric current density $J_i\lr{x} = \bar{\psi}\lr{x} \, \gamma_i \, \psi\lr{x}$ to the applied homogeneous electric field $E_j\lr{t}$ (where $t$ is the real Minkowski time). It is convenient to introduce the AC conductivity $\sigma_{ij}\lr{w}$, so that $\tilde{J}_i\lr{w} = \sigma_{ij}\lr{w} \, \tilde{E}_j\lr{w}$, where $\tilde{J}_i\lr{w} = \int dt \, e^{-i w t} J_i\lr{t}$ and $\tilde{E}_j\lr{w} = \int dt \, e^{-i w t} E_j\lr{t}$. Due to rotational symmetry of effective field theory (\ref{partfan}), the $\sigma_{ij}\lr{w}$ should have the form $\sigma_{ij}\lr{w} = \delta_{ij} \, \sigma\lr{w}$. Correspondingly, the DC conductivity is equal to the value of $\sigma\lr{w}$ at $w \rightarrow 0$.

 By virtue of the Green-Kubo dispersion relations \cite{Kadanoff:63:1, Asakawa:01:1, Aarts:07:1}, the Euclidean current-current correlators
\begin{eqnarray}
\label{corr}
 G\lr{\tau} = \frac{1}{2} \, \sum\limits_{i=1,2} \, \int dx^1 \, dx^2 \, \langle J_i\lr{0} \, J_i\lr{x} \rangle
\end{eqnarray}
can be expressed in terms of the $\sigma\lr{w}$ as
\begin{eqnarray}
\label{corr_eq}
 G\lr{\tau} = \int\limits^{\infty}_{0}
 \frac{dw}{2 \pi} \, K\lr{w, \tau} \, \sigma\lr{w} ,
\end{eqnarray}
where the thermal kernel $K\lr{w, \tau}$ is
\begin{eqnarray}
\label{kernel}
 K\lr{w, \tau} = \frac{w \cosh\lr{w \lr{\tau - \frac{1}{2T}} }}{\sinh\lr{\frac{w}{2T}}}
\end{eqnarray}
and $\tau \equiv x^0$ is the Euclidean time. We use here a nonstandard definition of the kernel (\ref{kernel}) from \cite{Aarts:07:1}, which is more convenient for numerical analysis.

 Note that the current density in graphene is the charge which flows through the unit length in unit time and thus has the dimensionality of $L^{-2}$ (where $L$ stands for length) in units with $\hbar = c = 1$. Correspondingly, the current density in lattice units is the charge which flows through a link of the dual lattice of length $a$ in time $a/v_F$. Thus in order to express the current-current correlator (\ref{corr}) in physical units, one should multiply the results obtained on the lattice by $a^2 \, v_F^2/a^4$, where the additional factor $a^2$ comes from integration over $x^1$, $x^2$ in (\ref{corr}). With the Euclidean time $\tau$ in (\ref{corr}), (\ref{corr_eq}) and (\ref{kernel}) being expressed in units of lattice spacing in temporal direction $a/v_F$, integration over $w$ in (\ref{corr_eq}) also includes a factor $v_F^2/a^2$. We thus conclude that the AC conductivity $\sigma\lr{w}$ is dimensionless. Moreover, the DC conductivity $\sigma\lr{0}$ is a universal quantity which does not depend on the lattice spacing or on the ratio of lattice spacings in temporal and spatial directions. For conversion to the SI system of units, it should be multiplied by $e^2/\lr{2 \pi h}$.

 In numerical simulations $G\lr{\tau}$ is measured for several ($\sim 10^1$) discrete values of $\tau$. A commonly used method to invert relation (\ref{corr_eq})
  and to extract the continuum function $\sigma\lr{w}$ from the lattice discretization of $G\lr{\tau}$ is the Maximum Entropy Method \cite{Asakawa:01:1, Aarts:07:1}. But this method suffers from some technical difficulties, so we use middle point of Euclidean current-current correlator to measure the conductivity at low frequencies (see \cite{Buividovich:2012nx} for further details):
   \begin{eqnarray}
\label{m_point}
G\lr{\frac{1}{2T}} \approx  \pi T^2 \sigma\lr{w}.
\end{eqnarray}

 For staggered fermions the electric current $J_i\lr{y}$ can be expressed in terms of the fields $\Psi_x$ as \cite{DeGrandDeTarLQCD}:
\begin{eqnarray}
\label{j_staggered}
 J_i\lr{y}  =
 \frac{1}{8} \sum\limits_{\eta} \,
 \delta_{\eta_3, \, 0} \, \delta_{\eta_i, \, 0} \,
 \left(
 \bar{\Psi}_{2 y + \eta} \alpha_{\eta, \, i} \Psi_{2 y + \eta + \hat{i}}
 + \right. \nonumber \\ \left. +
 \bar{\Psi}_{2 y + \eta + \hat{i}}
 \alpha_{\eta, \, i} \Psi_{2 y + \eta}
 \right)  ,
\end{eqnarray}
where we took into account that the spatial link variables $\theta_{x, i}$ are effectively equal to zero. Since current (\ref{j_staggered}) is defined on the lattice
 with double lattice spacing, we calculate the Euclidean current-current correlator (\ref{corr}) only on time slices with even $\tau$.

To study insulator-semimetal phase transition  it is useful  to consider the fermion condensate $ \vev{ \bar{\psi} \, \psi }$.
In the insulator phase $ \vev{ \bar{\psi} \, \psi } \neq 0$ and in the semimetal phase $ \vev{ \bar{\psi} \, \psi }=0$.
So, the fermion condensate $ \vev{ \bar{\psi} \, \psi }$ is the order parameter for the insulator-semimetal phase transition.
  In terms of staggered fermions it can be written as
\begin{eqnarray}
\label{condensate_def}
 \vev{ \bar{\psi} \, \psi }
 =
 \frac{1}{8 \, L_0 \, L_1 \, L_2} \, \sum\limits_{x, t} \, \vev{\bar{\Psi}_x \Psi_x }  .
\end{eqnarray}
 Since the fermions in the partition function are integrated out, the current-current correlator (\ref{corr}) and chiral condensate (\ref{condensate_def})
  can be expressed in terms of expectation values of certain combinations of the staggered fermion propagator $D^{-1}_{x, y}\lrs{\theta_{x, \mu}}$ with respect to weight    (\ref{ferm_integrated}).

\section{Numerical results and discussion}
\label{sec:num_results}

\begin{figure}[ht]
 \includegraphics[width =9.5cm]{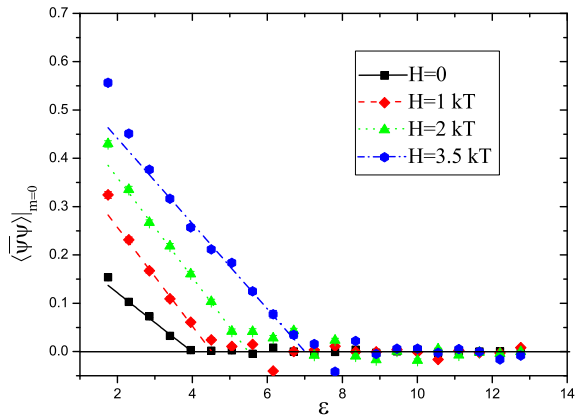}\\
 \caption{The fermion condensate $\vev{\bar{\psi} \, \psi}$ as a function of the dielectric permittivity of substrate $\epsilon$ in the
 limit $m \to 0$ for different values of $H$. Solid lines are the fits of the  data with the function
 $\vev{\bar{\psi} \, \psi} \sim \lr{\epsilon_c - \epsilon}^\gamma$ at different values of magnetic field.}
 \label{fig:condensates}
\end{figure}

 Using the algorithm described in the previous section, we have generated $100$ statistically independent gauge field configurations on
the $20^4$ lattice for each set of the parameters ( $\epsilon$, $m$, $H$ ). The results in the limit of zero mass was obtained by performing
simulations at several nonzero values of the $m=0.01,~ 0.02,~ 0.03$  and extrapolating the expectation values of physical observables
to $m \rightarrow 0$ for each pair ( $\epsilon$, $H$ ). The dielectric permittivity of the substrate
was varied within the range  $\epsilon = 1.75 \ldots 12.75$.  The magnetic field $H$ was varied in the interval which
corresponds 1--7 quanta of the magnetic flux  (\ref{quantum}) through the whole lattice. It should be noted that for the lattice
used in  the calculation one quantum of magnetic flux corresponds to the field $H_0=500$ T. This large value of the magnetic
field results from the fact that  $H_0 \sim 1/L_s^2$, $L_s$--is the size of graphene sample (spatial size of the lattice).
Although today our computers don't allow to consider lattice of the size much larger than $20^4$, we plan
to do this in future.

\begin{figure}[ht]
 \includegraphics[width = 9.5cm ]{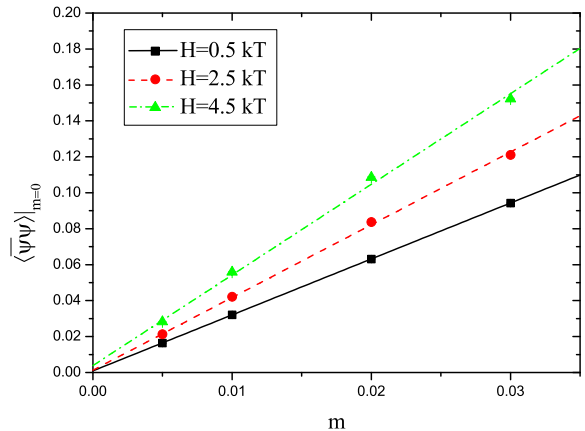}\\
 \caption{The fermion condensate $\vev{\bar{\psi} \, \psi}$ at dielectric permittivity $\epsilon=15$
 as a function of the fermion mass $m$ for the fields $H=0.5, 2.5, 4.5$ kT. Solid lines are the linear
  extrapolations to the massless limit.}
 \label{fig:conden_weak}
\end{figure}

In Figure \ref{fig:condensates} the fermion condensate $\vev{\bar{\psi} \, \psi}$ as a function of the dielectric permittivity of substrate
$\epsilon$ in the limit $m \to 0$ at different values of $H$ is shown. From this plot one sees that for each value of magnetic field
the condensate $\vev{\bar{\psi} \, \psi} \neq 0$  for small values of the $\epsilon$ and $\vev{\bar{\psi} \, \psi} = 0$  for large
values of the $\epsilon$. To determine the critical value $\epsilon_c$ the data for $\epsilon<\epsilon_c$ were fitted by the function
$\vev{\bar{\psi} \, \psi} =b \lr{\epsilon_c - \epsilon}^\gamma$ for each magnetic field $H$. Thus the critical permittivity $\epsilon_c$  as
a function of magnetic field was determined.

\begin{figure}[ht]
 \includegraphics[width = 9.5cm]{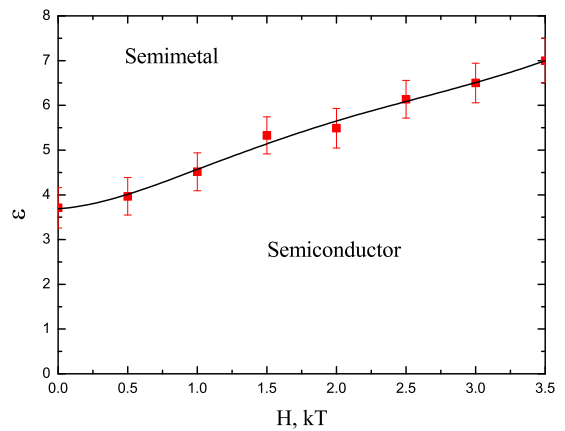}\\
 \caption{The phase diagram for graphene in the $(H, \epsilon)$ plane.}
 \label{fig:diagram}
\end{figure}

From the study of the fermion condensate $\vev{\bar{\psi} \, \psi}$ one can state that external magnetic field shifts
the insulator-semimetal phase transition in graphene to the direction of larger values of the $\epsilon$. 

 Physical interpretation of the observed phenomenon is the following. External magnetic field enhances
the density of the eigenfunctions of Dirac operator $D_{x, y}\lrs{\theta_{x, \, \mu}}$ at zero eivenvalue,
what leads to the enhancement of the condensate $\vev{\bar{\psi} \, \psi}$. The larger the value of
the condensate, the larger the value of the  $\epsilon$ is needed to screen the charges of quasiparticles
and destroy the condensate. Similar phenomenon takes place in QCD \cite{Buividovich:2008wf}.

In the paper \cite{Cea:2012up} the planar (2+1) quantum electrodynamics  with two degenerate (2+1) staggered fermions in an external magnetic
field  was investigated. The authors shown that external magnetic field led to dynamical mass generation
and appearance of nonzero fermion condensate $\vev{\bar{\psi} \, \psi} \neq 0$  in  weak coupling region.

Contrary to \cite{Cea:2012up} in this paper 3+1 electrodynamics was simulated. Within the accuracy of the calculation dynamical mass generation in the
weak coupling region has not been found. To study this question in more detail we have generated $400$ statistically
independent gauge field configurations on the $20^4$ lattice deep in the  weak coupling region ( $\epsilon=15$ ) for
different magnetic fields in the range $0.5-4.5$ kT. In Fig. \ref{fig:conden_weak} the fermion condensate $\vev{\bar{\psi} \, \psi}$
as a function of the fermion mass $m$ for different values of $H$ is shown. The extrapolation of the data
to massless limit shows that within the uncertainty of the calculation the condensate $\vev{\bar{\psi} \, \psi}=0$
for all magnetic fields under consideration.
This result confirms that there is no dynamical mass generation in 3+1 electrodynamics with 2+1 massless fermions.

The study of the fermion condensate which was done above allows us to draw the phase diagram for graphene
in the $(H, \epsilon)$  plane  shown in Fig \ref{fig:diagram}.

\begin{figure}[ht]
 \includegraphics[width = 9.5cm]{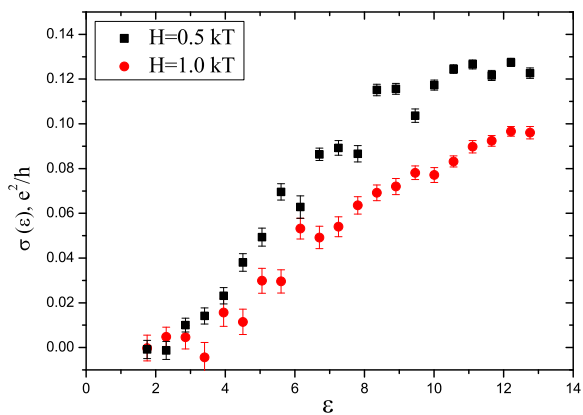}\\
 \caption{The conductivity $\sigma$ as a function of the $\epsilon$ for the fields $H=0.5,~1$ kT. }
 \label{fig:conductivity_a}
\end{figure}

\begin{figure}[ht]
 \includegraphics[width = 9.5cm]{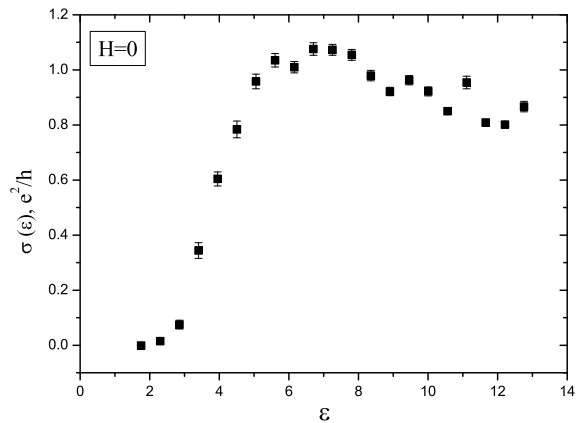}\\
 \caption{The conductivity $\sigma$ as a function of the $\epsilon$ without magnetic field.}
 \label{fig:conductivity_b}
\end{figure}

  The measurement of the conductivity of graphene can be done through the calculation of the current-current correlators (\ref{corr}). The details of the calculation can be found in paper \cite{Buividovich:2012uk}.
In Fig. \ref{fig:conductivity_a} the conductivity $\sigma$ for the fields $H=0.5,~1$ kT as a function of the $\epsilon$ is presented.
To compare the conductivity in external magnetic field with the ones without magnetic field we plot Fig. \ref{fig:conductivity_b}.
From this plots one sees that in external magnetic field the insulator-semimetal phase transition becomes broader.
The value of the conductivity in magnetic field in the weak coupling region is by an order of magnitude smaller
than that without magnetic field. The last fact can be explained by the possible appearance of the another fermion condensate which cannot be adequately described by the staggered fermions due the lack of the full U(4) symmetry of the continuous effective field model in the lattice simulations.

In conclusion, in this paper the results of numerical simulation of monolayer graphene in external magnetic field is presented.
The numerical simulations are performed in the effective lattice field theory with noncompact $3 + 1$-dimensional
Abelian lattice gauge fields and $2 + 1$-dimensional staggered lattice fermions. The dependence of the fermion
condensate and graphene conductivity on the dielectric permittivity of substrate for different values of external
magnetic field is calculated. It is found that magnetic field shifts the insulator-semimetal phase transition
to larger values of the dielectric permittivity of substrate. The phase diagram of graphene in external
magnetic field is drawn.

\begin{acknowledgments}
 The authors are much obliged to Dr. Timo Lahde who was the first to draw their attention to  graphene field theory. The authors are grateful to Prof. Mikhail Zubkov for interesting and useful discussions. The work was supported by Grant RFBR-11-02-01227-a and by the Russian Ministry of Science and Education, under contract No. 07.514.12.4028. Also the work was supported by the FEFU development program project "Nuclear Medicine". Numerical calculations were performed at the  ITEP
system Stakan (authors are much obliged to A.V. Barylov, A.A. Golubev, V.A. Kolosov, I.E. Korolko, M.M. Sokolov for the help), the MVS 100K at Moscow Joint Supercomputer Center and at Supercomputing Center of the Moscow State University.
\end{acknowledgments}


\end{document}